\documentclass[twocolumn,prb,aps,superscriptaddress,showpacs,amsmath,amssymb,floatfix]{revtex4-1}
\usepackage{bm}
\usepackage{epsfig}
\usepackage{graphicx}

\begin{document}

\title{Joint density of states and its relationship with quasiparticle interference pattern in $d$-wave
superconductors}

\author{Qiang Han}
\affiliation{Department of Physics, Renmin University of China,
Beijing, China}
\affiliation{Department of Physics and Center of
Theoretical and Computational Physics, The University of Hong Kong,
Pokfulam Road, Hong Kong, China}

\author{Dan-Bo Zhang}
\affiliation{Department of Physics, Renmin University of China,
Beijing, China}

\author{Z. D. Wang}
\affiliation{Department of Physics and Center of Theoretical and
Computational Physics, The University of Hong Kong, Pokfulam Road,
Hong Kong, China}

\date{\today}
\begin{abstract}
We present analytic analyses of the joint density of states
(JDOS) and the elastic scattering susceptibility (ESS) for $d$-wave
superconductors under the linear dispersion approximation. The JDOS and ESS diverge on the
same curves which are found to be the envelope functions
related to the contour of constant energy of the Bogoliubov quasiparticles.
The relative amplitudes of the divergence on the envelope functions are derived.
It is also found that certain octet vectors come close to the vertices of the
envelope curves when the Dirac-cone anisotropy is strong, which is relevant to the $d$-wave cuprates.

\end{abstract}

\maketitle

The scanning tunneling spectroscopy(STS) has long been applied to
exploration of the local single-particle spectrum of high-$T_c$
cuprates ~\cite{Fischer}. Recently, modulations of the local density
of states (LDOS) have been studied by the Fourier transformed (FT)
STS and analyzed within the quasiparticle interference (QPI) picture
that attributes the QPI pattern to the elastic Bogoliubov quasiparticle
scattering from impurities ~\cite{Hoffman,McElroy03nature}.  The
characteristic wave vectors seen in the FT-LDOS are interpreted by
an effective octet model \cite{McElroy03nature,qhwang} that proposes
seven wave vectors connecting the tips of the banana-shaped contours
of constant quasiparticle energy (CCE).  Numerical calculations
using the $T$-matrix formalism
\cite{qhwang,scalapino,franz,csting,Zhu} on the scattering from a
weak single impurity were in agreement with the octet model.

As is known, the angle resolved photoemission spectroscopy (ARPES)  and the STS
are complementary to each other, both measuring the single-particle
electronic structure with the former being in the momentum space and the latter in
the real one. A more intimate relation between them has been noted
recently by comparing the two-particle information extracted from
these two measurements~\cite{McElroy}, namely the susceptibility of
the superconductor to impurity embodied by the FT-LDOS intensity and
the joint density of states by evaluating the autocorrelation
of the spectral function observed by ARPES. A remarkable matching between the peak positions in
the autocorrelated ARPES data and the octet vectors implies likely that the JDOS has an intrinsic relationship with
the QPI patterns.
However, the inherent physical reason of the above-mentioned matching is unclear. Moreover, to the best of our knowledge, there still lacks a
simple theoretical relation between the FT-LDOS, which is closely
related to the imaginary part of the autocorrelation of the
Nambu-Gor'kov Green's function  in the weak scattering limit, and
the JDOS.

%Within the T-matrix formalism  and weak coupling limit the FTLDOS are encoded in a two-particle
%quantity that closely related to the autocorrelation of the Gor'kov Green's function.
%and two-particle information, namely the susceptibility of the superconductor to
%scatterers in the charge, spin and/or pairing channels, has been extracted.
%One apparent common feature possessed by the scanning tunneling spectroscopy
%(STS) and the angle resolved photoemission spectroscopy (ARPES)  is that they
%both probe the generalized single-particle density of states (DOS) in real
%space (i.e. local DOS) or in momentum space (i.e. the spectral function).
%Information of the two-particle spectra can be extracted by Fourier transform
%(FT) of the STS data, or by evaluating the autocorrelation of the ARPES spectra, and after comparing them
%more intimate relation between these two probes was found recently.~\cite{McElroy}

In this paper, we attempt to uncover the inherent physics of the above-mentioned
intimate relation between the JDOS and the QPI
pattern. % observed recently by comparing the ARPES data to the FT-STS results.
In particular, an envelope scenario is proposed to analyze the divergent behavior of the JDOS,
which reveals its highly structured pattern. The matching of the octet wave vectors
and the JDOS peak positions is also investigated.

Let us first derive the basic formulas for both the FT-STS and the
JDOS as well as their intrinsic relationship. The %elastic scattering susceptibility
ESS for the weak nonmagnetic impurity associated with the FT-STS is
defined as
\begin{equation}
S(\mathbf{p},\omega) = -\frac{1}{\pi}\text{Im}\sum_\mathbf{k} [G(\mathbf{k+p},\omega^+) \sigma_3 G(\mathbf{k},\omega^+)]_{11},
\end{equation}
where $\omega^+\equiv\omega+i0^+$ and $G$ is the Nambu-Gor'kov Green's function of unperturbed $d$-wave superconductors,
\begin{equation}
G^{-1}(\mathbf{k},z) =\left(
\begin{array} {cc}
G_{11} & G_{12} \\
G_{21} & G_{22}
\end{array}
\right)^{-1} = \left(
\begin{array} {cc}
z-\xi_\mathbf{k} & -\Delta_\mathbf{k} \\
-\Delta_\mathbf{k} & z+\xi_\mathbf{k}
\end{array}
\right),
\end{equation}
with $\xi_\mathbf{k}=\sum_{\bm \delta} t_{\bm \delta}
e^{i\mathbf{k}\cdot{\bm \delta}}-\mu$ the normal-state  dispersion
measured with respect to the chemical potential,
$\Delta_\mathbf{k}=-2\Delta_0(\cos k_x - \cos k_y)$ the $d$-wave gap
function and
$E_\mathbf{k}=\sqrt{\xi_\mathbf{k}^2+\Delta_\mathbf{k}^2}$ the
Bogoliubov quasiparticle energy. To facilitate the derivation, we
define two correlation functions $L$ and $M$:
\begin{equation}
\begin{aligned}
    L_\omega(\mathbf{p},z) &=& \sum_{s=\pm,\mathbf{k}} G_{11}(\mathbf{k}+s\mathbf{p},z) A(\mathbf{k},\omega), \\
    M_\omega(\mathbf{p},z) &=& \sum_{s=\pm,\mathbf{k}} G_{12}(\mathbf{k}+s\mathbf{p},z) B(\mathbf{k},\omega),
\end{aligned}
\label{LM}
\end{equation}
where $A(\mathbf{k},\omega)=-\text{Im}
G_{11}(\mathbf{k},\omega^+)/\pi$ is the single-particle spectral
function and $B(\mathbf{k},\omega) =
-\text{Im}G_{21}(\mathbf{k},\omega^+)/\pi$. It is straightforward to find that
the ESS can be expressed as
\begin{equation}
    S(\mathbf{p},\omega) = \text{Re} [L_\omega(\mathbf{p},\omega^+) - M_\omega(\mathbf{p},\omega^+)]. \label{formS}
\end{equation}
On the other hand, the JDOS %, which is related to the autocorrelation of the ARPES data,
is given by
\begin{equation}
    J(\mathbf{p},\omega) =\sum_\mathbf{k} A(\mathbf{k+p},\omega)
    A(\mathbf{k},\omega), \label{acfunc}
\end{equation}
and can simply be  expressed in terms of $L_\omega$:
\begin{equation}
    J(\mathbf{p},\omega) = -\frac{1}{2\pi} \text{Im}
    L_\omega(\mathbf{p},\omega^+). \label{formJ}
\end{equation}
By examining Eqs.~(\ref{formS}) and (\ref{formJ}), one can readily
find that $S(\mathbf{p},\omega)$ and $J(\mathbf{p},\omega)$ are
respectively related to the real and imaginary parts of the correlation function
$L_\omega(\mathbf{p},\omega^+)$. Therefore
they show singular behaviors on the common curves/surfaces in general, which stem from the interplay of the poles
of the Green's function and spectral function in Eq.~(\ref{LM}).
This argument is verified in the following detailed discussions.

\begin{figure}[ht]
\begin{tabular}{cc}
\epsfig{figure=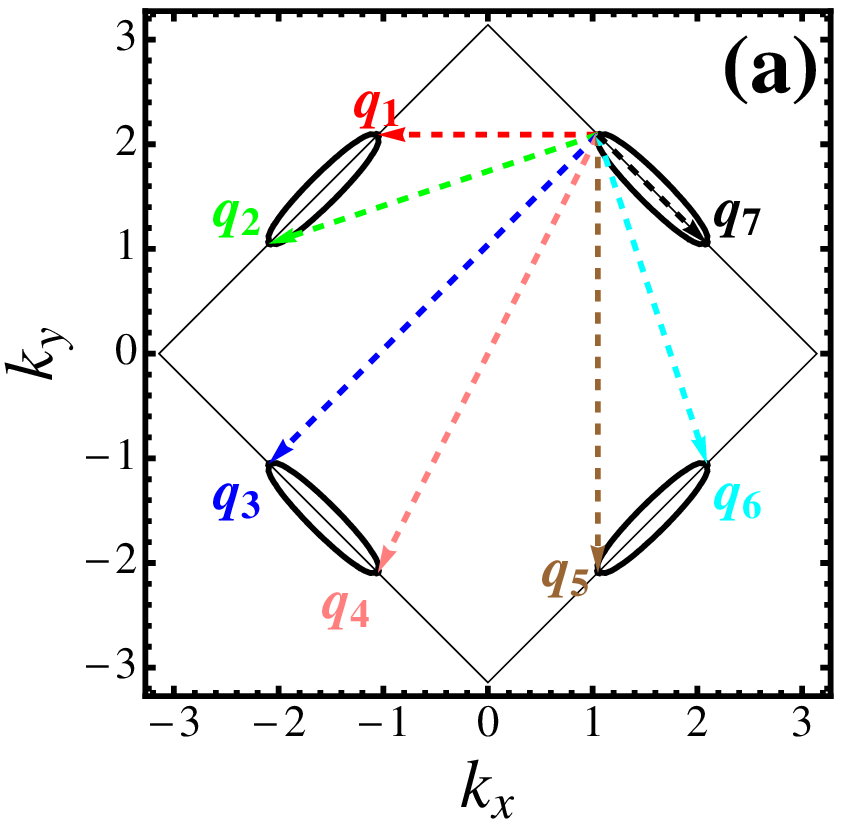,width=0.20\textwidth} & \epsfig{figure=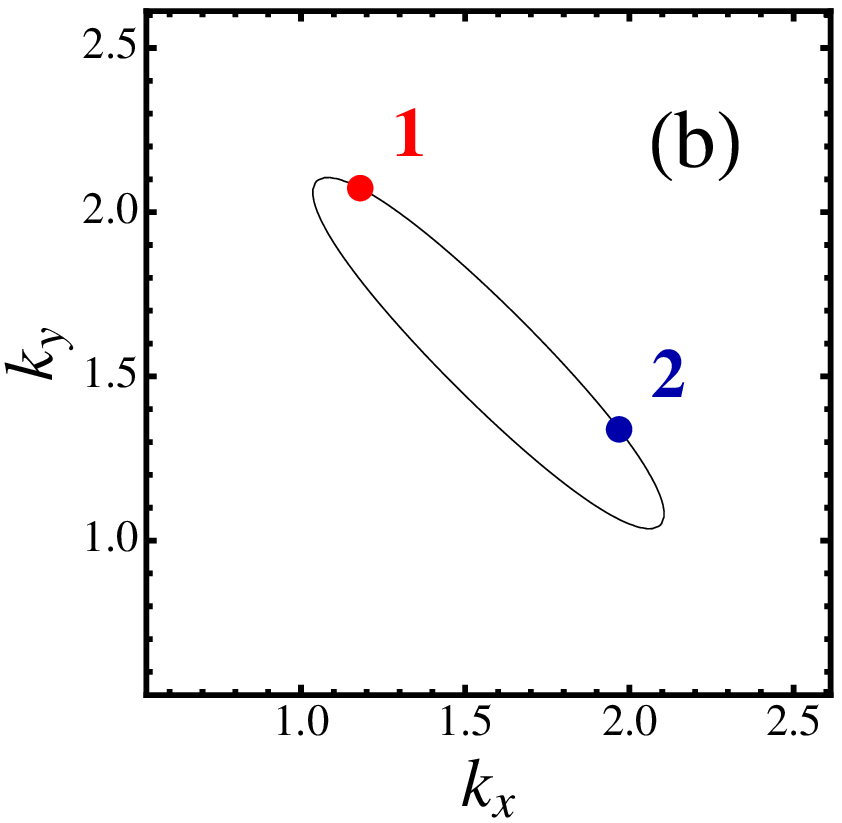,width=0.20\textwidth} \\
\epsfig{figure=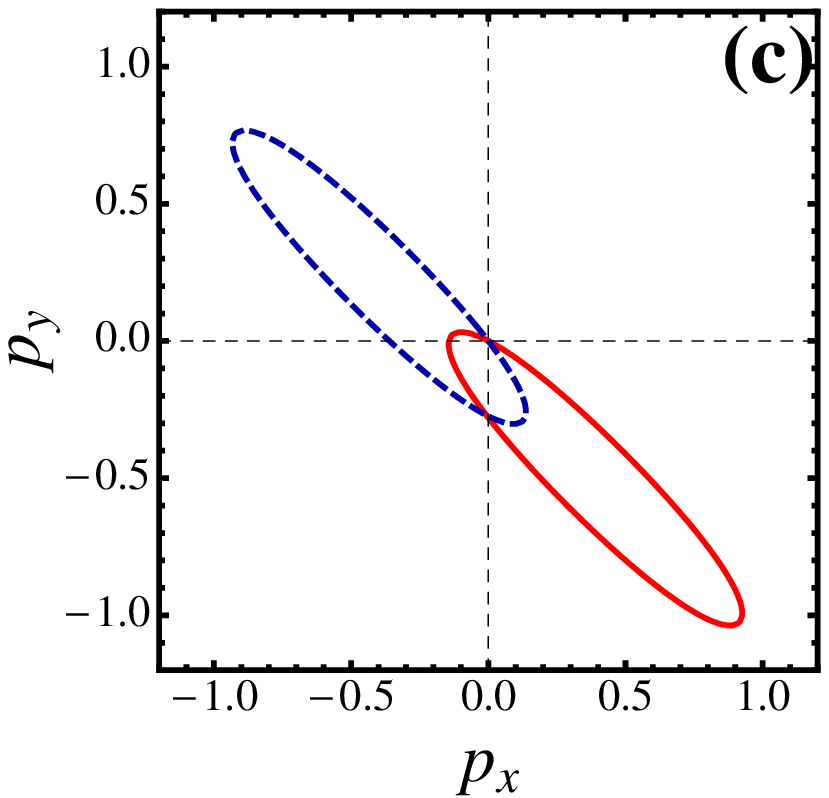,width=0.20\textwidth} & \epsfig{figure=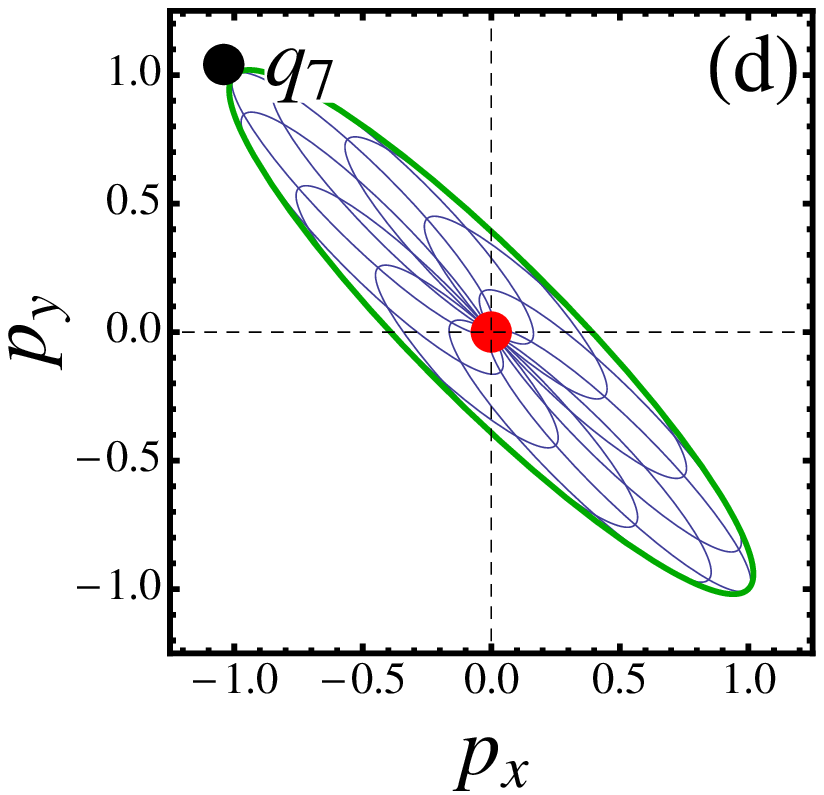,width=0.20\textwidth} \\
\end{tabular}
\caption{Illustration of the idea of the envelope model.
% Here for simplicity we use a tight-binding model with only non-zero nearest-neighbor hopping integral $t$ and the chemical potential is set as zero.
Here for simplicity we choose $t^\prime=\mu=0$ and
$\Delta_0=0.2t$. (a) CCE's determined by
$E_\mathbf{k}=|\omega|$ with $\omega=-0.4t$. The normal state Fermi
surface is also shown. (b), CCE-1 in the first quadrant of FBZ.
Two $\mathbf{k}$ points  (red and blue dots) are
shown. (c), two curves determined by $E_\mathbf{k+p}=|\omega|$ with red solid corresponds
to $\mathbf{k}$ point 1 while blue dashed one to 2 in
(b). (d), the envelope is shown in green solid line and central
red dot. For schematic illustration we also draw in blue color a
cluster of curves for several different
$\mathbf{k}$'s. $\mathbf{q}_7$ as shown in (a) is located on the tip
of the elliptic envelope. 
%$J(\mathbf{p},\omega)$ has inverse square root divergence on the envelope with relative amplitude $(p_x-p_y)^2$, and accordingly $\mathbf{q}_7$ has the largest JDOS.
} \label{fig1}
\end{figure}

To look into the ESS and JDOS of $d$-wave superconductors more
specifically, we now attempt to derive the explicit formulae of
$L_\omega(\mathbf{p},z)$ and $M_\omega(\mathbf{p},z)$. The spectral
functions $A$ and $B$ in Eq.~(\ref{LM}) restrict that the momenta
$\mathbf{k}$'s, which contribute to $L_\omega$ and $M_\omega$, are
determined by the four banana-shaped CCE's given by
$E_\mathbf{k}=|\omega|$. Taking into account the $d$-wave
structure of the gap, we here use
the Dirac cone approximation around the four gap nodes
$\mathbf{n}_{1,2,3,4}=(\pm\eta,\pm\eta)$ where the index of each node is the same as the quadrant it locates.
For instance, near the first node
$\mathbf{n}_1=(\eta,\eta)$, $\Delta_\mathbf{k}\approx v_\Delta k_1$,
$\xi_\mathbf{k}\approx v_\text{F} k_2$ and $E_\mathbf{k}\approx
\sqrt{(v_\Delta k_1)^2+(v_\text{F} k_2)^2}$, where
$k_{1,2}=(\mathbf{k}-\mathbf{n}_1)\cdot(\mathbf{e}_x\mp\mathbf{e}_y)/\sqrt{2}$
in the local coordinate system of the first node. The $\hat{k}_1$
($\hat{k}_2$) axis is along the direction of increasing
$\Delta_\mathbf{k}$ ($\xi_\mathbf{k}$). If we perform the integration
over $\mathbf{k}$ in Eq.~(\ref{LM}) around the four gap nodes,
analytical results can be obtained for the three cases: 1) $\mathbf{p}$
is small such that $\mathbf{k}$ and $\mathbf{k+p}$ sit at the
vicinity of the same node, with $\mathbf{q}_7$ depicted in
Fig.~\ref{fig1}(a) being belong to this case; 2) $\mathbf{p}-(
\mathbf{n}_3-\mathbf{n}_1)$ is small such that $\mathbf{k}$ and
$\mathbf{k+p}$ sit around the first and third nodes respectively,
with $\mathbf{q}_{3,4}$ being belong to this case; 3) $\mathbf{p}-(
\mathbf{n}_2-\mathbf{n}_1)$ is small such that $\mathbf{k}$ and
$\mathbf{k+p}$ sit around the first and second nodes respectively,
with $\mathbf{q}_{1,2,5,6}$ being belong to this case.

For the first case, after tedious derivations, we obtain
\begin{equation}
    L_\omega(\mathbf{p},\omega^+) = - \frac{1}{2\pi v_\Delta
    v_\text{F}} f_\omega^+ (\mathbf{p},\omega^+), \label{formL}
\end{equation}
\begin{equation}
    M_\omega(\mathbf{p},\omega^+) = - \frac{1}{2\pi v_\Delta
    v_\text{F}} f_\omega^- (\mathbf{p},\omega^+), \label{formM}
\end{equation}
\begin{equation}
    f_\omega^\pm
    (\mathbf{p},z) = \frac{\omega^2-\left(E_\mathbf{p}/2\right)^2
    \pm \omega^2 \cos^2\theta}{E_\mathbf{p}\sqrt{E_\mathbf{p}^2-4z^2}} +
    \frac{1}{4}, \label{fpm}
\end{equation}
where the angular factor $\cos\theta\equiv v_\Delta
p_1/E_\mathbf{p}$. Substituting Eqs.~(\ref{formL}) and (\ref{formM}) into
Eqs.~(\ref{formS}) and (\ref{formJ}), we have
\begin{equation}
    S(\mathbf{p},\omega) = \left\{
    \begin{array}{l}
    -\frac{1}{\pi v_\Delta v_\text{F}}
    \frac{\omega^2 \cos^2\theta}{E_\mathbf{p} \sqrt{E_\mathbf{p}^2-4\omega^2}},\ \ \
    2|\omega| < E_\mathbf{p}, \\
    0,\ \ \ 2|\omega| > E_\mathbf{p},
    \end{array}
    \right.
\label{Sexact}
\end{equation}
and
\begin{equation}
    J(\mathbf{p},\omega) = \left\{
    \begin{array}{l}
    0,\ \ \ 2|\omega| < E_\mathbf{p}, \\
    \frac{1}{4\pi^2 v_\Delta v_\text{F}}
    \frac{\omega^2-\left(E_\mathbf{p}/2\right)^2 + \omega^2
    \cos^2\theta}{E_\mathbf{p}\sqrt{4\omega^2-E_\mathbf{p}^2}}, \ \
    \ 2|\omega| > E_\mathbf{p}.
    \end{array}
    \right.
\label{Jexact}
\end{equation}
It can be seen clearly that both the ESS and JDOS show
inverse-square-root divergence when approaching the same characteristic curve
governed by $E_\mathbf{p}=2|\omega|$, which will be termed as {\it
envelope} in the following. The singular parts of both
functions originate from the same complex function as shown in
Eq.~(\ref{fpm}). The ESS is nonzero(zero) outside(inside) the
envelope while the JDOS is just the opposite. The vertices of this elliptic envelope are just the
octet vectors $\pm\mathbf{q}_7$ corresponding to $\theta=0,\pi$, on which both the ESS and JDOS reach
the largest amplitude of divergence. This angular dependence of the ESS stems from the
coherent factors of the $d$-wave superconductors as found in Ref.~[\onlinecite{franz}] and here
we find that the JDOS has the same angular dependence.

For the second case,
$\mathbf{p}^\prime=\mathbf{p}-(\mathbf{n}_3-\mathbf{n}_1)$ is small.
We find that the ESS and  JDOS have the same analytic
expressions as Eqs.~(\ref{Sexact}) and (\ref{Jexact}) if the corresponding numerators are respectively replaced by
$ \omega \sin\theta^\prime
(\omega\sin\theta^\prime-E_{\mathbf{p}^\prime}/2)$
and  $(\omega \sin\theta^\prime -
E_{\mathbf{p}^\prime}/2)^2$,  where
$\sin\theta^\prime\equiv v_\text{F}p^\prime_2/E_{\mathbf{p}^\prime}$. Therefore both quantities diverge on
the envelope $E_{\mathbf{p}^\prime}=2|\omega|$. However the angular behaviors of the ESS and JDOS are distinct from
each other and from their behaviors in the first case. Accordingly both the ESS and JDOS do not reach their largest values on the vertices of the elliptic envelope i.e.~$\pm\mathbf{q}_4$ ($\theta^\prime=0,\pi$). Another octet vector $\mathbf{q}_3$ locates on the envelope center where the JDOS has $\delta$-function divergence, in agreement with the ARPES data \cite{McElroy}. However the ESS
vanishes at $\mathbf{q}_3$, seemingly inconsistent with the FT-STS observation. The reason lies in the fact that the ESS reaches its strongest divergence at a vector of the envelope with $\theta^\prime=-\text{sgn}(\omega)\pi/2$, which is closest to $\mathbf{q}_3$ and therefore corresponds to the observed QPI peak as expected.

For the third case with $\mathbf{p}$ near $\mathbf{n}_2-\mathbf{n}_1$,
we find again that both the ESS and JDOS diverge on certain common curves, although the explicit formulas for
both quantities are rather involved than the above two cases.
In order to capture the main physics without lengthy derivations, we propose an envelope scenario
in the following to first locate these curves where both the quantities exhibit divergent behaviors and then obtain the relative amplitudes of the divergence.

To introduce the {\it envelope scenario}, we examine the JDOS in detail, which is relatively easier to handle.
The spectral function of pure $d$-wave superconductors can be expressed as
\begin{equation}
A(\mathbf{k},\omega) = u_\mathbf{k}^2 \delta(\omega-E_\mathbf{k}) +  v_\mathbf{k}^2 \delta(\omega+E_\mathbf{k}), \label{specfunc}
\end{equation}
where the coherence factors are
$u_\mathbf{k}^2=\frac{1}{2}(1+\frac{\xi_\mathbf{k}}{E_\mathbf{k}})$ and $v_\mathbf{k}^2=\frac{1}{2}(1-\frac{\xi_\mathbf{k}}{E_\mathbf{k}})$.
Here we only study the $\omega<0$ case for comparison with the autocorrelated ARPES data \cite{McElroy}.
Therefore, we have after combining Eqs.~(\ref{acfunc}) and (\ref{specfunc}) for $\omega<0$,
\begin{equation}
    J(\mathbf{p},\omega) = \int v_\mathbf{k+p}^2 v_\mathbf{k}^2 \delta(\omega+E_\mathbf{k+p}) \delta(\omega+E_\mathbf{k}) \frac{d^2\mathbf{k}}{(2\pi)^{2}}.
     \label{acnegative}
\end{equation}
Leaving aside the coherence factors, one can easily find that the
$\mathbf{k}$ points in the first Brillouin zone (FBZ), which give
the largest contribution to the integral, are determined from the two CCE
governed by the two $\delta$ functions.

In Fig.~\ref{fig1}, we show the CCEs given by
$E_\mathbf{k}=|\omega|$, which consists of 4 banana-shaped closed
curves and are denoted as CCE-1,2,3,4 hereafter. To better elucidate
our idea, we first focus on the intra-CCE contribution (i.e. both
$\mathbf{k}$ and $\mathbf{k+p}$ on the same banana) to the JDOS. We
examine the CCE in the first quarter of the FBZ, i.e. CCE-1.
Corresponding to each $\mathbf{k}$ on CCE-1, all $\mathbf{p}$ points
satisfying $E_\mathbf{k+p}=|\omega|$ form a closed curve denoted as
$U_\mathbf{k}(\mathbf{p})$, which has exactly the same shape as
CCE-1 except that the center is shifted as shown in
Fig.~\ref{fig1}(c). Therefore, for all $\mathbf{k}$'s on CCE-1,
$U_\mathbf{k}(\mathbf{p})$'s form a one-parameter family of curves
whose envelope is plotted in Fig.~\ref{fig1}(d) shown as the green line and
red dot. One can readily see that the envelope gives the locus on
which the JDOS is most likely strong because it represents the
position where curves in the family touch or intersect most.

Quantitative analysis can be obtained under the Dirac cone approximation.
Accordingly CCE-1 is an ellipse whose semi-axis lengths are
$a\equiv|\omega|/v_\Delta$ and $b\equiv|\omega|/v_\text{F}$. The
corresponding family of curves $U_\mathbf{k}(\mathbf{p})$'s as shown
in Fig.~\ref{fig1}(d) are determined from the following equations, $
(p_1,p_2) =(a\cos\theta-k_1, b\sin\theta-k_2)$ with $(k_1,k_2) =
(a\cos\phi,b\sin\phi)$, where $\theta, \phi \in [0,2\pi)$. The
envelope of the above family is composed of an ellipse and a
trivial isolated point,
\begin{equation}
E_\mathbf{p}=2|\omega|, \ \ \ E_\mathbf{p}=0. \label{env1}
\end{equation}
Taking advantage of the envelope functions, we can
evaluate the JDOS's divergence behavior just on these curves by
integrating Eq.~(\ref{acnegative}), which is relatively simpler
than calculating $J(\mathbf{p},\omega)$ for arbitrary $\mathbf{p}$.
We find that the relative amplitude of the inverse-square-root divergence is
proportional to $\cos^2\theta$ on the elliptic envelope and the
divergence is a $\delta$ function on the isolated point.
As for the inter-CCE case between CCE-1 and CCE-3, the envelope consists of an ellipse and an isolated point similar to the intra-CCE case
except that the center is shifted by a vector
$\mathbf{n}_3-\mathbf{n}_1$ and the relative amplitude on the ellipse changes to $(1+\sin\theta)^2$.
All these results are in a full agreement with former explicit formulas.

\begin{figure}[ht]
\epsfig{figure=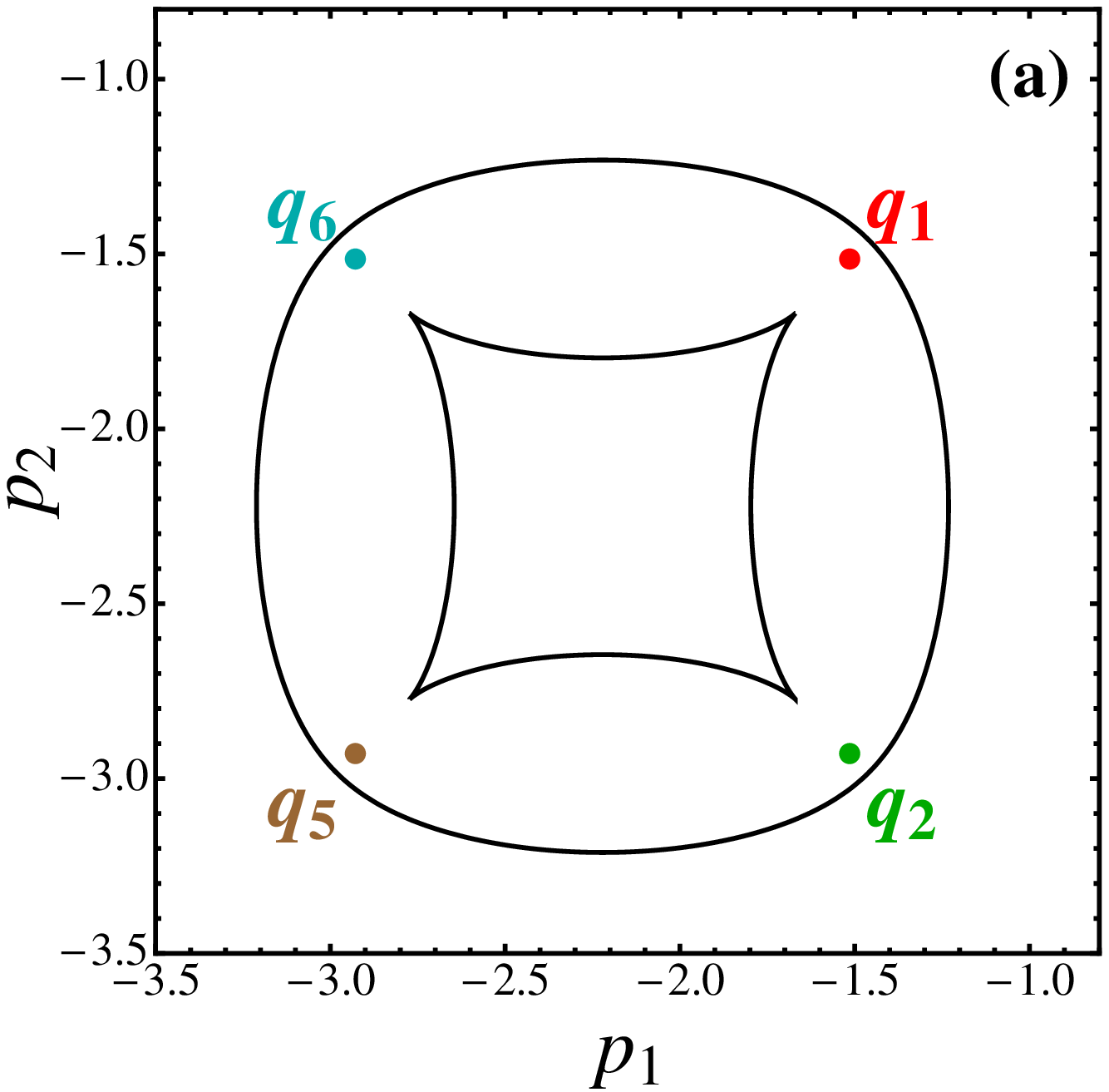,width=0.22\textwidth}
\epsfig{figure=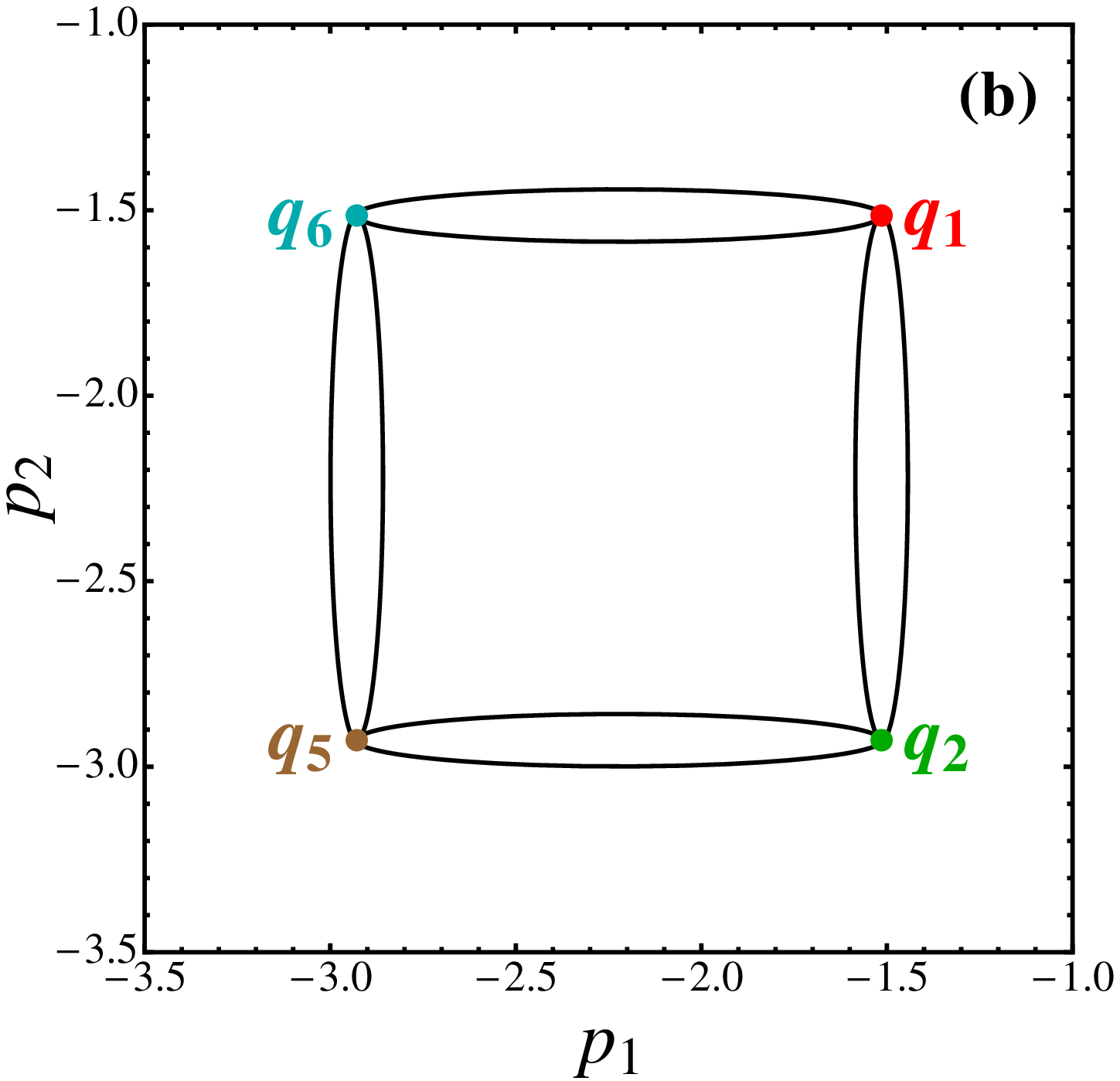,width=0.22\textwidth}
\caption{Plot of the envelope function according to Eq.~(\ref{U12}), as well as 4 octet momenta $\mathbf{q}_{1,2,5,6}$. (a), less anisotropic case $\gamma=v_\Delta/v_\text{F}=0.4$ ($\Delta_0=0.4t$, $\omega=-0.8t$). (b) strong anisotropic case $\gamma=v_\Delta/v_\text{F}=0.1$ ($\Delta_0=0.1t$, $\omega=-0.2t$). Note that the anisotropy is determined by $\Delta_0/t$. We choose different $\omega$ just to have similar octet $q$'s for this two cases.
}
\label{fig2}
\end{figure}
More importantly, the envelope analysis can be applied to the inter-CCE case between CCE-1 and CCE-2,
which is otherwise rather complicated. The family
of curves $U_\mathbf{k}(\mathbf{p})$'s satisfy,
$(p_1,p_2)=-(k_\text{F},k_\text{F})+(b\cos\theta-k_1,a\sin\theta -
k_2)$, where $k_\text{F}=\sqrt{2}\eta$ denotes the length of the
Fermi vector at the nodes.
%\begin{equation}
%\left\{
%\begin{array}{cc}
%p_1 = -k_\text{N} + b\cos\theta - k_1, \\
%p_2 = -k_\text{N} + a\sin\theta - k_2,
%end{array}
%\right.
%\end{equation}
The corresponding envelope is
\begin{equation}
\left\{
\begin{array}{cc}
    p_1 = -k_\text{F} + a\cos\alpha \left[ \gamma^2 s_\gamma(\alpha) \pm r_\gamma(\alpha) \right], \\ %
    p_2 = -k_\text{F} + a\sin\alpha  \left[ s_\gamma(\alpha) \pm \gamma^2 r_\gamma(\alpha) \right], %
\end{array}
\right.
\label{U12}
\end{equation}
where $r_\gamma(\alpha)=(\cos^2\alpha +
\gamma^2\sin^2\alpha)^{-1/2}$ and $s_\gamma(\alpha)=(\sin^2\alpha +
\gamma^2\cos^2\alpha)^{-1/2}$ with $\gamma\equiv
b/a=v_\Delta/v_\text{F}$ is the anisotropic factor. This envelope is
plotted in Fig.~\ref{fig2}, where a circle-like ["$+$" sign in
Eq.~(\ref{U12})] and an astroid-like curve ["$-$" sign in Eq.~(\ref{U12})]
centered at $(-k_\text{F},-k_\text{F})$\cite{note} is clearly seen.  Also shown
are the 4 wave vectors ($\mathbf{q}_{1,2,5,6}$ as in Fig.~\ref{fig1}(a))  according to the
octet model. One of our main findings is that, the
anisotropy of the quasiparticle excitation spectrum plays a
significant role in bridging the JDOS picture and the octet model.
We find that the octet vectors  $\mathbf{q}_{1,2,5,6}$ approach the cusps (most divergent as shown later) of the
astroid for sufficient large anisotropy of the Dirac cone, while depart from the cusps of the astroid for the less
anisotropic case, as shown in Fig.~\ref{fig2}. This indicates that
the effectiveness of the octet model\cite{Hoffman,McElroy03nature,qhwang} is actually
associated with the strong anisotropy of the quasiparticle excitations
in $d$-wave cuprates.

\begin{figure}[ht]
\epsfig{figure=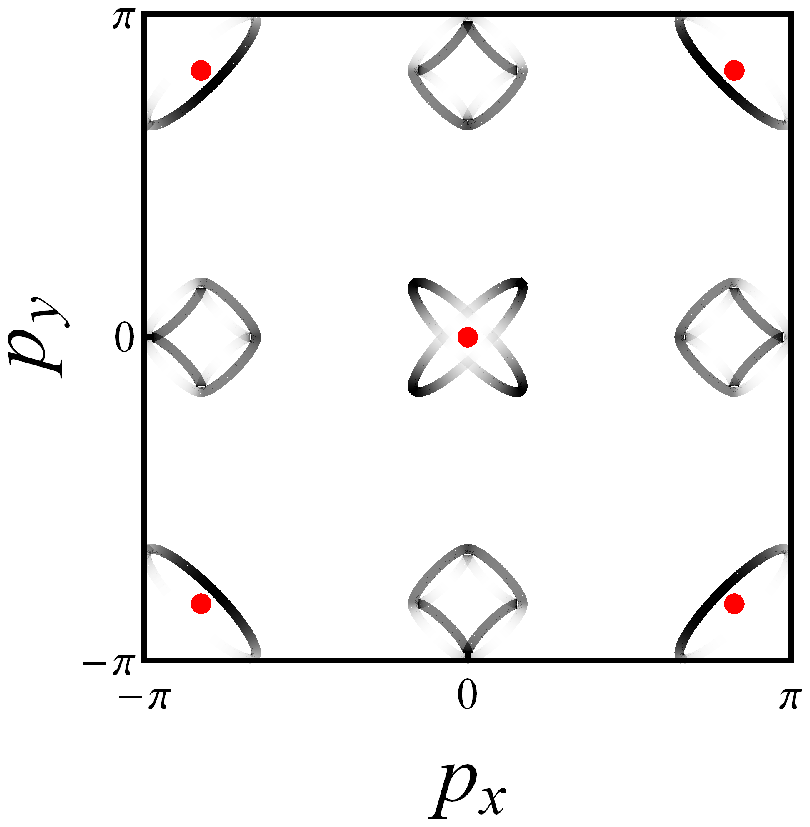,height=0.22\textwidth}
\epsfig{figure=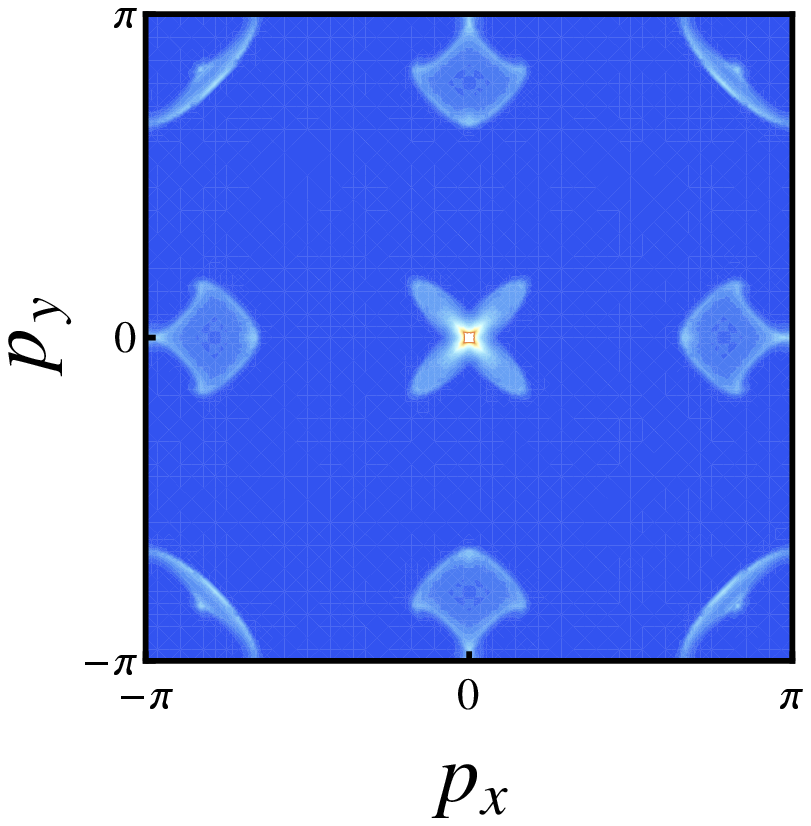,height=0.22\textwidth}
\caption{
Comparison of analytical and numerical results of the JDOS in the first Brillouin zone.
The parameters are $t^\prime=-0.3 t$, $\mu=-1.0 t$, $\Delta_0=0.2t$, $\omega=-0.2t$.
Left: Plot of all the envelopes. %, including the intra-CCE, inter-CCE and all the equivalent procedures allowed by symmetries.
The gray levels on the curves signal the relative amplitudes of divergence.
%Red dots denote the centers of all the ellipses.
Right:
%JDOS calculated by numerical integration
Numerical results according to Eq.~(\ref{acnegative}) with lattice size $400\times 400$, where $\delta(x)\approx\frac{\eta_0}{\pi(x^2+\eta_0^2)}$ with $\eta_0=0.01t$.
}
\label{fig3}
\end{figure}

On  the circle-like and astroid-like envelopes, we again find the inverse
square root divergence of JDOS similar to the intra-CCE case, with the
relative amplitude (the coefficient of the divergent terms),
\begin{equation}
F_\pm(\alpha)  \propto g_\pm(\alpha) \left[ \left|
\frac{s_\gamma(\alpha)}{r_\gamma(\alpha)} \pm
\frac{r^2_\gamma(\alpha)}{s^2_\gamma(\alpha)}
\right|^{-\frac{1}{2}}+ (r\leftrightarrow s)\right], \label{Jinter}
\end{equation}
with
\begin{equation}
g_\pm(\alpha) = [1 \pm \gamma r_\gamma(\alpha) \sin\alpha ][1+\gamma
s_\gamma(\alpha)\cos\alpha]. \label{num}
\end{equation}
According to Eq.~(\ref{Jinter}), the amplitude $F_{-}(\alpha)$ on the astroid
diverges when approaching its 4 cusps at $\alpha=\pi/4,3\pi/4,5\pi/4,7\pi/4$, coinciding with
the octet vectors $\mathbf{q}_{1,2,5,6}$ in the large anisotropic case as shown in Fig.~\ref{fig2}.
This observation agrees well with the autocorrelation of the ARPES data \cite{McElroy}. The effect of
the coherence factors is embodied by $g_\pm(\alpha)$ in
Eq.~(\ref{Jinter}), inducing an additional angular dependence of the amplitudes. Accordingly the amplitudes on
the astroid- and circle-like envelopes are highly uneven with one half is rather larger
than the other half, breaking the four-fold symmetry possessed by the envelopes.
Thus these two pronounced parts combine together to form a duck-foot structure as shown in Fig.~\ref{fig3}.
Note that in the same spirit, the relative amplitude of divergence of the ESS on these envelopes can also be obtained
(not shown here as the quantitative numerical analyses of the ESS have been presented before\cite{qhwang,scalapino,franz,csting,Zhu}).

Combining the above analyses, we can plot all the envelopes where
$J(\mathbf{p},\omega)$ diverges in Fig.~\ref{fig3}(a) with the
relative amplitude of divergence being taken into consideration. We
find two cigars centered at the origin, four duck feet on the
coordinate axes and four (semi)cigars on the zone corners as well as five
red dots. The envelope analysis capture the essential structures and
symmetries of the JDOS, in excellent agreement with the
experimental data \cite{McElroy} as well as the JDOS by numerical
integration of Eq.~(\ref{acnegative}) as shown in
Fig.~\ref{fig3}(b). As for the peak positions, $\mathbf{q}_7$ and
$\mathbf{q}_4$ locate on vertices of the central and corner cigars
respectively. $\mathbf{q}_{1,2,5,6}$ is on apexes of the duck feet.
$\mathbf{q}_3$ lies on the center of the corner cigars
denoted as red dot. All these features are in full consistent with
the experimental observations~\cite{McElroy}.

In summary, the JDOS and ESS in $d$-wave superconductors
have been investigated analytically, both being found to
diverge on the curves of envelopes. Using the envelope scenario by us,
we have shown clearly that the the octet vectors locate
on the vertices of the envelopes validating the octet model for the $d$-wave cuprates with strong anisotropy of
the quasiparticle excitation spectrum.
The relative amplitude of the divergence of the JDOS has been obtained.
Finally, we wish to pinpoint that the envelope scenario may also be used to
investigate the JDOS of the iron-based superconductors,
which may shed light on the Fermiology and pairing symmetry in this new material.

%Similar analysis can also be applied to the studying
%of the quasiparticle interference pattern of the iron-based
%superconductors.

The work was supported by the RGC of Hong Kong under Grant
No.HKU7055/09P and HKUST3/CRF/09, the URC fund of HKU, the Natural
Science Foundation of China No.10674179.


\begin{thebibliography}{99}
\bibitem{note}
In the local coordinate system the center of the astroid- and circle-like envelopes locates at $(k_\text{F},k_\text{F})$. More generally, it is $\mathbf{n}_2 - \mathbf{n}_1$.
\end{thebibliography}
\end{document}